\documentclass{article}

\usepackage[english]{babel}

\usepackage[letterpaper,top=2cm,bottom=2cm,left=3cm,right=3cm,marginparwidth=1.75cm]{geometry}

\usepackage{amsmath}
\usepackage{graphicx}
\usepackage{graphicx}
\usepackage{subcaption}
\usepackage[colorlinks=true, allcolors=blue]{hyperref}
\usepackage{orcidlink}
\usepackage[authoryear, round]{natbib}

\title{Reflections on the Future of Statistics Education in a Technological Era}
\author{
    Craig Alexander\textsuperscript{1,*,\textdagger} \orcidlink{0000-0001-6734-747X},
    Jennifer Gaskell\textsuperscript{1} \orcidlink{0000-0001-9583-323X},  
    Vinny Davies\textsuperscript{1} \orcidlink{0000-0003-1896-8936} \vspace{0.5cm} \\ 
    \textsuperscript{1}School of Mathematics and Statistics, University of Glasgow \\
    \textsuperscript{*}Corresponding author email - Craig.Alexander.2@Glasgow.ac.uk \\
    \textsuperscript{\textdagger}All authors contributed equally to this work. 
}
\date{}

\begin{document}
\maketitle

\begin{abstract}
Keeping pace with rapidly evolving technology is a key challenge in teaching statistics. To equip students with essential skills for the modern workplace, educators must integrate relevant technologies into the statistical curriculum where possible. University-level statistics education has experienced substantial technological change, particularly in the tools and practices that underpin teaching and learning. Statistical programming has become central to many courses, with R widely used and Python increasingly incorporated into statistics and data analytics programmes. Additionally, coding practices, database management, and machine learning now feature within some statistics curricula. Looking ahead, we anticipate a growing emphasis on artificial intelligence (AI), particularly the pedagogical implications of generative AI tools such as ChatGPT. In this article, we explore these technological developments and discuss strategies for their integration into contemporary statistics education.
\end{abstract}

\section{Introduction}

The way we teach is constantly evolving, and this is especially true in the statistical sciences. Advances in computing power and data storage have significantly expanded our ability to collect and analyse data. As methods and models continue to develop to accommodate increasingly large and complex datasets, the discipline is shifting towards approaches that were not traditionally considered part of statistics, necessitating a deeper understanding of computational thinking and data-driven workflows. Broadly categorised under `data science', this field encompasses Statistics, Machine Learning (ML), and Artificial Intelligence (AI), with the distinctions between them often blurred \citep{diggle2015statistics,ij18}. The rapid expansion of these techniques raises important questions about how best to integrate them into university-level statistical curricula \citep{hardin2015data}.

Statistics education has always adapted to changes in technology and practice. While many core statistical principles remain largely unchanged over long periods, other parts of the curriculum have evolved considerably. Earlier shifts were strongly linked to the move towards modern programming languages, particularly with the adoption of R for statistical computing. More recently, however, these changes have extended beyond programming languages themselves. New data types, evolving coding practices, and increasingly complex modelling approaches are reshaping how statistics is applied in both research and industry. In this article, we consider how these technological developments are influencing the statistics curriculum and how educators might respond to them within existing programmes. We do not focus on educational technologies such as live polling tools, as, while important, they are not specific to statistical education and have been explored in detail elsewhere \citep{bond20,granic22}.

We begin by discussing how programming practices have evolved, including the rise of R, the development of tidyverse workflows, and the ongoing debate around teaching base R, tidyverse, or a combination of both. We then consider how Python might be introduced alongside R, including the potential benefits of multi-language teaching and the implications for student cognitive load. The paper then moves to changes in data and coding practice, exploring how large or unstructured data sources, accessed through application programming interfaces (APIs) and web scraping, are influencing statistical work. We discuss how these topics might be integrated gradually across programmes rather than taught in isolation, alongside the growing role of version control and reproducible workflows as important skills for modern statisticians.

Next, we examine how ML and AI may be incorporated into statistics curricula. We argue that the depth and extent of coverage should depend on graduate pathways and programme goals. In some cases, ML can be integrated into existing modules, while in others dedicated courses may be appropriate. More advanced AI content is likely to depend on the availability of educators with relevant expertise. Finally, we consider the rapid emergence of generative AI tools, discussing different attitudes to their use in teaching, how students are already engaging with them, and the implications for assessment, marking, and feedback. Throughout, we reflect on how these developments shape teaching practice and the challenges they present for educators.

This article is structured as follows. Section~\ref{sec:R_Python} examines developments in statistical programming, including the role of RStudio  and the integration of Python within statistics curricula \citep{kenett2022modern}. Section~\ref{sec:data_code} discusses modern data sources, strategies for teaching data management, and best practices for sharing and storing code efficiently. Section~\ref{sec:ML_AI} explores the integration of ML and AI into statistical education, and Section~\ref{sec:genAI} assesses the impact of generative AI tools such as ChatGPT on statistical learning and assessment.


\section{Programming within the Statistics Curriculum}
\label{sec:R_Python}

In recent decades, advances in computing and the development of programming languages in the field of statistics have led to a transformation in the way the modern statistician carries out analyses.  Traditional statistical methods, once confined to manual calculations and limited software tools, have evolved into complex data-driven approaches supported by programming environments designed for efficiency, scalability, and reproducibility. Languages such as R, Python, and Julia have become essential for statisticians, offering extensive libraries for data analysis, visualisation, and ML. The development of such programming languages has helped to shape modern statistical practices, enabling researchers and analysts to process large datasets and develop innovative methodologies that were once considered impractical.

The development of statistical software began in the mid-20th century, when early computing machines were first used to automate statistical calculations. In the 1960s and 1970s, languages like FORTRAN were implemented to create some of the earliest bespoke statistical programs, such as SAS. Following these first examples of statistical software, an increase of specialised statistical software such as SPSS and S-PLUS became available to statisticians, expanding upon earlier software, providing a more user-friendly interface for data analysis, and expanding the suite of modelling techniques available to the user. 

The emergence of R in 2000 revolutionised statistical computing by offering an open-source, extensible platform for statisticians. As free software, its accessibility and reach quickly surpassed many competitors, which often required a licence. As a command-line program, R provided greater flexibility in the range and scale of tasks it could perform, supported by a substantial ecosystem of external packages. This flexibility gave it a clear advantage over menu-driven alternatives such as Minitab and SPSS. However, the shift towards R also required many educators to transition from more familiar software environments to a scripting-based language, presenting both technical and pedagogical challenges as staff adapted their teaching materials and approaches.

As R became established within academic and professional practice, it increasingly assumed a central role in statistics education. Among the statistical software packages available, it has emerged as a widely used language in modern curricula. As the demand for modelling skills using software has grown, the question in many statistics programmes is no longer whether R should be taught, but how it should be integrated into the curriculum to best prepare students for the workplace. While R remains central to contemporary statistical computing, its place in education continues to evolve. Many academic institutions now embed R programming across statistics modules, and its use has expanded beyond traditional disciplinary boundaries. In the following sections, we examine how the teaching of R has developed within higher education, considering its pedagogical value, approaches to curriculum design, and, subsequently, how Python can be integrated alongside R to support a broader computational skillset.

\subsection{R in the Statistics Curriculum}
\label{sec:dev_R}

As the use of R has grown across academia and industry, particularly alongside increased attention to `big data’ problems, the tools and facilities available to R users have expanded significantly. This demand has had a transformative effect on statistics education, with R programming now commonplace in many statistics curricula. In this context, many users choose to work with R through an integrated development environment (IDE), which provides a more supportive interface for writing, running, and managing code. One of the most commonly used IDE's for R is RStudio.

RStudio extends the functionality of base R by providing a graphical user interface (GUI), that is, a visual interface through which users interact with code and data, alongside features such as syntax highlighting, auto-completion of code, and integrated data visualisation tools. These features help reduce the learning curve for beginners and support good coding practice. The open-source nature of RStudio ensures compatibility across major operating systems and promotes accessibility by removing financial barriers, allowing students to install and use the software freely on their own devices.

Presenting statistical software in an accessible and user-friendly manner is important for shaping students’ experiences and attitudes towards statistics. Considerations of usability have therefore motivated many recent developments within both R and RStudio, several of which are discussed in the following sections. While these tools can significantly enhance teaching and learning, their effective integration into the curriculum depends on educators possessing, or developing, the relevant expertise to use them confidently and pedagogically effectively.

\subsubsection{Tidyverse}
One of the most commonly used packages within R is the tidyverse suite of packages \citep{tidyverse19}. The underlying philosophy of this collection is to promote the practice of tidy data \citep{wickham14}, encouraging users to structure data in a consistent format from the outset to support a more efficient analysis workflow \citep{wickham23}. The tidyverse itself comprises multiple packages designed to support the different stages of a statistical analysis, from data import and transformation to visualisation and modelling. Its popularity has also contributed to increased use of R in disciplines beyond statistics, including psychology \citep{ryan21}, social science \citep{imai22}, and medicine \citep{musa23}.

The pedagogical advantages over base R commands stem from the unified design philosophy of tidyverse, which prioritises consistency, clarity and simplifies concepts for beginners. Unlike base R, where functions can have inconsistent naming conventions and require nested syntax for complex operations, tidyverse offers a more coherent set of packages with clear naming conventions for function (for example, \texttt{filter()}, \texttt{mutate()}), helping to reduce the cognitive load for learners. Central to the tidyverse suite is the use of the pipe operator, which replaces base R's nested function calls with step-by-step workflows. For example, an operation which transforms variables in data may require several nested functions using base R becomes a logical sequence of verbs in tidyverse, reflecting how learners naturally think through problems. This not only improves code readability but encourages modular problem solving, breaking analyses into smaller, interpretable steps \citep{johnson97}.

Visualisation is also simplified through tidyverse by use of the ggplot package. Though base R plotting functions (e.g. \texttt{plot()}, \texttt{hist()}) are flexible, they lack the layered grammar-of-graphics framework of ggplot. By building plots incrementally by adding aesthetics, geometries and themes, learners develop a structured understanding of data visualisation similar to constructing sentences in language \citep{wickham11}.


Although there are clear pedagogical benefits to teaching R using tidyverse, the question arises whether we should teach only tidyverse functionality over base R, or vice versa, or a hybrid of both. With a tidyverse only approach, the learning curve is lower which is ideal for beginners and allows them to construct readable code for stepwise thinking. There is also a relevance within industry of knowing tidyverse, as this is commonly used within modern industry roles. The downside to this approach is learners will effectively skip base R commands having a limited low-level understanding of R and potentially avoiding fundamental programming concepts and core operators within R. This can also cause issues for using certain statistical modelling methods, which use base R conventions. 

Teaching with base R only is somewhat the `standard' approach across institutions, providing a solid foundational knowledge of R and core programming principles such as loops and vector operations. Though more of a `forced' skill, beginning with base R can teach good debugging skills while working through some of the quirks of R. The main downside to using only base R is the steeper learning curve for learners, particularly those with no programming experience and inconsistent syntax. 

Studies in the literature suggest that students can achieve similarly positive learning experiences when the same concepts are taught through different syntactic approaches, provided that the teaching is carefully structured \citep{carscadden22}. In addition, \cite{cetinkaya21} highlights how the grammar and design consistency of tidyverse can support understanding across the data analysis cycle. A hybrid approach to teaching R, first introducing foundational programming concepts through base R before incorporating tidyverse tools for data analysis, can therefore be a reasonable strategy. Such an approach can foster a deeper understanding of tidyverse methods by grounding them in foundational knowledge of base R. The principal challenge lies in allocating sufficient curriculum time to cover both perspectives at a pace that does not overwhelm learners. When carefully structured, however, a hybrid model can also offer flexibility within teaching teams, enabling educators with strong tidyverse expertise to lead tidyverse-based components, while others may teach using base R or tidyverse as appropriate to the module content and their own experience.

\subsubsection{Shiny}
The creation of RStudio has also lead to the development of effective communication libraries and tools for reporting data analysis to a non-technical audience. The shiny package \citep{shiny} allows for the simple development of web applications in RStudio, providing an approach for constructing dynamic web-based applications, using R, avoiding the need for learners to require knowledge in HTML or JavaScript.

A common theme within the workplace for a statistician is expanding beyond conducting analysis and incorporating effective communication. Shiny has emerged as a key tool for bridging this gap between theory and application and has been promoted within higher education as effective teaching tools \citep{potter16,fawcett18}. Shiny allows for the transformation of methodology into user-friendly applications, allowing students to explore data dynamically, visualise patterns within model frameworks, and develop understanding by building apps to communicate methods, allowing the learner to critically think about the method in a deeper manner. 

The use of shiny within assessments can aid critical thinking when structuring apps, building functional applications which reinforce key programming skills. New concepts such as reactive programming are also explored through shiny, developing skills in structuring code in a modular fashion. Including shiny within the statistics curriculum allows us to help bridge the gap between theory and application, promoting good practice in effective communication while further developing knowledge.

\subsubsection{RMarkdown \& Quarto}

The notions of reproducible research and effective data communication are fundamental to both scientific and industry practice. Tools such as RMarkdown \citep{rmarkdown} and Quarto \citep{quarto} support these principles and have therefore become increasingly common components of modern statistics curricula. Their inclusion reflects a broader shift in statistics education towards practices that mirror real-world analytical workflows, where analysis, documentation, and communication are closely intertwined.

The integration of R into the statistics curriculum has contributed to a move towards more project-based learning, with open-ended assignments that require students to carry out statistical analyses and produce written reports. Tools such as RMarkdown and Quarto encourage learners to combine code, narrative, and results within a single document, reinforcing good practice in reproducibility and transparent research. In doing so, they foreground the importance of not only conducting an analysis, but also clearly communicating its purpose, assumptions, and conclusions.

Including RMarkdown and Quarto within the statistics curriculum therefore supports the development of skills that extend beyond technical computation. By requiring students to structure their reasoning, justify modelling choices, and present results coherently, these tools help cultivate habits that are directly transferable to both academic research and the workplace. As a result, their use provides a natural and authentic mechanism for assessing statistical understanding alongside communication and reproducibility.

\subsection{Polyglot Programming}

The emphasis on reproducible workflows and integrated communication naturally leads to broader questions about the role of programming environments in statistics education, particularly as analytical practice increasingly spans multiple programming languages. Recent developments around Quarto reflect this shift and are closely linked to changes in the tools and services offered by Posit, the developers of RStudio.

One motivation for these developments has been to move beyond a solely R-focused ecosystem and to support the integration of additional languages such as Python and Julia. Through Quarto and recent extensions to RStudio, users can now work with multiple language engines within a single session, enabling genuinely cross-language workflows. This reflects common practice in industry, where analysts often adopt a pragmatic, task-driven approach to language choice rather than relying on a single programming language.

As mixed-language workflows become more common across academia and industry, there is a growing case for broadening the programming exposure provided within statistics curricula. While R remains central to statistical education, languages such as Python and Julia are increasingly visible in data science, computational modelling, and ML. Graduates are therefore likely to encounter professional environments in which multiple languages coexist, each selected according to context and task. Introducing students to this wider computational landscape can better prepare them for collaborative and interdisciplinary settings.

Expanding beyond a single language, however, is not straightforward. Teaching two programming languages can be challenging, particularly when many statistics educators were trained primarily in R and may have limited experience with Python, while familiarity with Julia is less common still. Although both Python and Julia offer strong technical capabilities, Python is currently more widely adopted across academia and industry, making it the more practical addition where curriculum space is constrained and introducing three languages would be unrealistic. This raises important questions about how best to introduce Python alongside R in statistics education, and whether the goal should be a unified environment or familiarity with multiple tools.

\subsubsection{Introducing Python}
\label{sec:Python}

Although Python predates R, having been introduced in 1991, its inclusion within statistics curricula has largely occurred in more recent years. This shift has been driven primarily by Python’s prominence in ML and data science, where widely adopted libraries such as scikit-learn \citep{pedregosa2011scikit} have made it a natural choice for teaching modern computational methods. As a result, Python has increasingly been positioned alongside R in statistics education, particularly in contexts that emphasise predictive modelling and ML.

While comparisons between R and Python have been explored in the literature \citep{Kalyan18, ozgur2017b}, the choice between languages in practice is often shaped less by technical differences and more by the surrounding ecosystem of packages and developer communities. Many core data structures and workflows have close similarities across the two languages, for example data frames in R and pandas \citep{pandas} in Python, or ggplot2 \citep{ggplot2} in R and visualisation frameworks such as matplotlib \citep{matplotlib} and its grammar-of-graphics inspired counterpart plotnine \citep{plotnine} in Python. Similarly, established statistical modelling frameworks developed in R, such as mgcv \citep{mgcv}, now have Python interfaces or related implementations, including pyGAM \citep{pyGAM} and pymgcv \citep{pymgcv}. While these translations support cross-language adoption, they often feel less natural than their native counterparts, reflecting differences in language design and typical usage patterns.

However, important differences remain, particularly in areas driven by machine learning and deep learning. Python has become the dominant interface for these methods, supported by mature and well-integrated frameworks such as TensorFlow \citep{tensorflow}, JAX \citep{jax}, and PyTorch \citep{pytorch}. While these frameworks are primarily accessed through Python, their computational backends rely heavily on optimised compiled code and hardware-accelerated libraries, enabling efficient large-scale and computationally intensive workflows. Combined with strong industry support and active development communities, this results in interfaces that are often more accessible and flexible for modern machine learning applications. By contrast, many advanced statistical methods continue to emerge first within R, reflecting the language’s close ties to the statistical research community.

These differences are also shaped by the backgrounds of the respective developer communities. R packages are frequently developed by statisticians and methodologists, often alongside new theoretical contributions, whereas Python libraries are more commonly produced by researchers and engineers working in ML, software development, and industry. This divergence influences both the design priorities of packages and the contexts in which each language is most naturally applied.

The growing prominence of ML within statistical research and education therefore provides a strong motivation for introducing Python within statistics curricula. While Python is unlikely to replace R as the sole language of instruction in the near term, there is an increasing case for exposing students to multiple programming languages. Such an approach reflects the diversity of modern analytical practice and acknowledges that different tools may be better suited to different tasks. This perspective motivates a closer examination of how R and Python might be taught together within statistics programmes, and what pedagogical models are best suited to supporting multi-language learning.

\subsubsection{Teaching both R and Python}
\label{sec:Combine_R_Python}

If both R and Python are to be taught within a degree programme, decisions about how they are introduced and integrated require careful consideration, particularly regarding curriculum progression and the order in which students encounter each language. While arguments can be made for introducing either language first, the practical reality for many statistics programmes is that R is already embedded across multiple modules. In such cases, introducing R early and reinforcing it throughout the programme, before later incorporating Python, is likely to be a coherent and pragmatic approach.

A further consideration is how Python should be incorporated into existing courses. In a course on linear regression, for example, it is common to provide example code and data to support problem-based learning. In this context, an inclusive approach is to provide equivalent Python code alongside existing R examples, rather than treating Python as an optional or peripheral addition. While presenting material in multiple languages risks increasing students’ cognitive load if not handled carefully, providing access to both languages can offer a valuable additional resource, allowing students to compare approaches and supporting future learning beyond the immediate course context.

\begin{figure}[t]
    \centering
    \begin{subfigure}[b]{0.8\textwidth}
        \includegraphics[width=\textwidth]{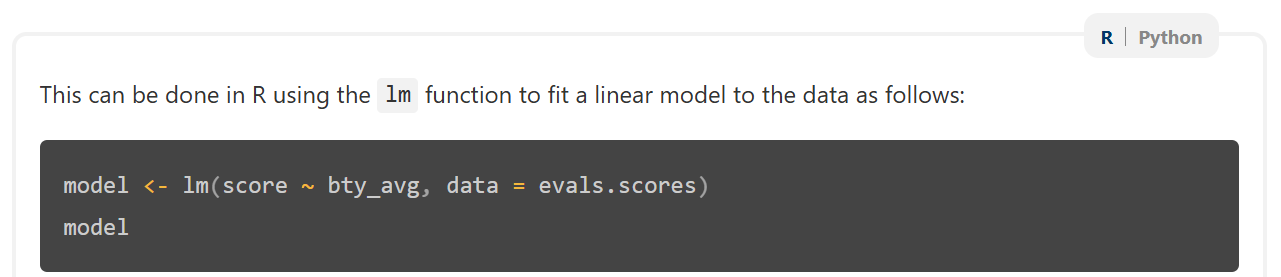}
        \caption{R code enabled}
        \label{fig:switcher_R}
    \end{subfigure}
    \begin{subfigure}[b]{0.8\textwidth}
        \includegraphics[width=\textwidth]{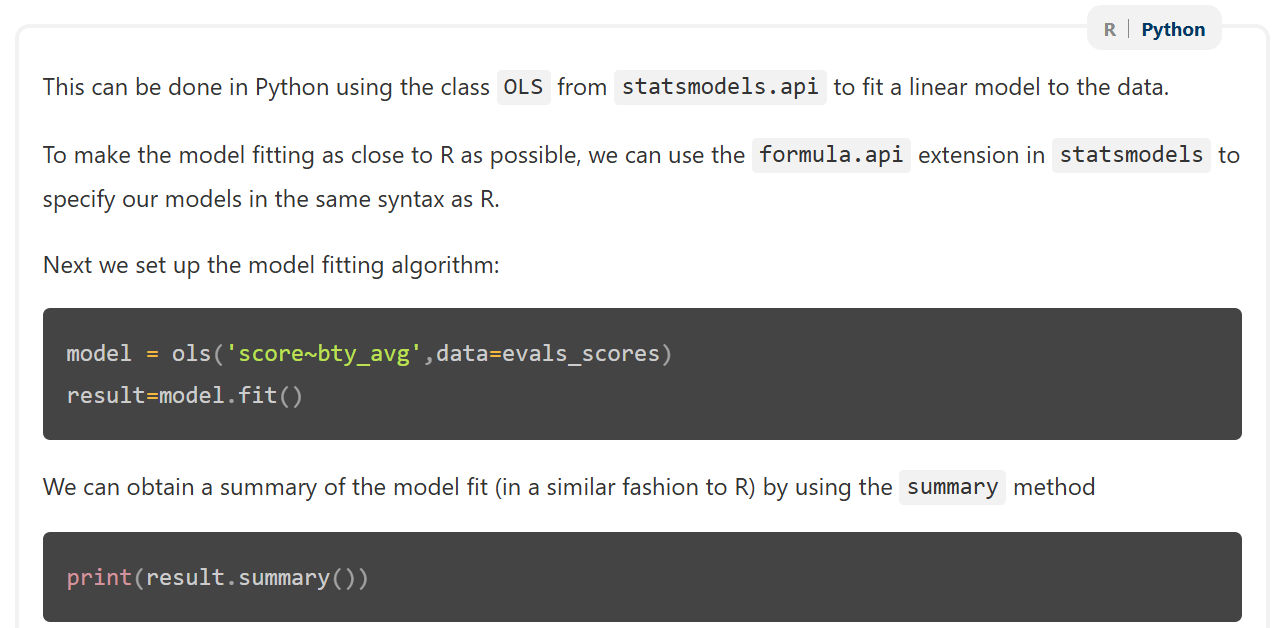}
        \caption{Python code enabled}
        \label{fig:switcher_Python}
    \end{subfigure}
\caption{An example of the language switcher, shown in the top right of each image. (a) R code enabled, allowing students to view the notes as if they were available only in R. (b) Equivalent view with Python enabled.}
    \label{fig:switcher}
\end{figure}

One approach to managing this cognitive load is the use of language switchers embedded within teaching materials, for the example the approach discussed in \cite{jack2023reflections}. An implementation of this type is illustrated in Figure~\ref{fig:switcher}, where students can toggle between R and Python code within a single set of notes, allowing them to focus primarily on one language while retaining access to the other as a reference or future learning resource. While the example tool is custom-built, similar functionality is now supported through tools such as Quarto, making it increasingly feasible for educators to provide comparable multi-language materials without relying on bespoke institutional solutions.

While offering materials in both languages provides flexibility for learners, it also introduces practical challenges in delivery. Many statistics educators may be less familiar with Python, just as ML or AI specialists may not necessarily know R. In such cases, a pragmatic approach may be to provide materials in both languages while teaching primarily in one, depending on staff expertise. This model requires appropriate support, both for developing materials in the alternative language and for ensuring that student questions can be addressed by someone with relevant experience. Shared teaching responsibilities or scheduled drop-in sessions may help meet this need. There will also be situations where equivalent implementations in both languages are not feasible. Even so, using both languages where possible remains beneficial, and openly explaining why certain components are language-specific can help students develop a more realistic understanding of mixed-language practice.

\section{The Influence of Data and Coding Practices}
\label{sec:data_code}

The growing prevalence of large-scale data, collected across many aspects of life and business, alongside advances in parallel computing and cloud infrastructure, has significantly influenced how statistical analysis is conducted. The scale and complexity of modern datasets increasingly require not only statistical expertise, but also familiarity with coding and computational tools. As these technologies continue to shape the field, they place greater emphasis on efficient data processing and scalable analytical methods, challenging some traditional statistical workflows.

Despite these developments, statistical education does not always fully reflect this evolving landscape. While the core principles of statistics are typically usually well covered, the integration of coding practices and computational techniques varies considerably across programmes. As a result, some students graduate with limited exposure to tools and methods for working with large datasets and distributed computing environments, both of which are becoming increasingly common in research and industry.

Addressing this gap may require rethinking aspects of how statistics is taught. Incorporating modern coding practices, such as code optimisation, version control, and the use of cloud-based resources, has the potential to better equip students for contemporary data challenges. By complementing traditional statistical theory with applied computational training, programmes can help students develop skills that are increasingly relevant in data-driven contexts.

This section examines how technological shifts have influenced statistical practice and explores how changes in teaching and assessment might better align statistical education with the demands of modern data analysis.

\subsection{Modern data sources \& structures}
\label{sec:data_origin}

With advances in technology and the widespread use of digital systems in both professional and everyday contexts, modern data sources are increasingly abundant and offer rich opportunities for use in statistics teaching. Such data may originate from websites, social media platforms, smart devices, open public datasets, Internet of Things (IoT) systems, and research infrastructures. While the range of available data is broad, accessing and working with these sources can present challenges. Modern datasets are often large, unstructured, or stored in formats that are not traditionally encountered in statistics education. This section therefore considers how modern data are commonly structured and highlights the computational considerations required to work with them effectively.

Most modern data can be broadly categorised as structured, semi-structured, or unstructured. Structured data remains the most common format used in statistics teaching and is typically organised into rigid schemas, such as tables with predefined fields. Large structured datasets are often stored in databases and accessed using SQL, which is already a familiar component of many statistics programmes. As a result, structured data continues to play a central role in teaching foundational statistical concepts.

Semi-structured data does not conform to strict tabular formats but contains identifiable elements, such as tags or metadata, that support processing and transformation. Common examples include XML and JSON files, which are frequently accessed through application programming interfaces (APIs). Unstructured data, by contrast, lacks a predefined format and often requires substantial pre-processing before analysis, with examples including free text, web-scraped content, and images. Introducing students to semi-structured and unstructured data offers opportunities to broaden the curriculum, for instance through the use of API queries to access public datasets or data generated through students’ own digital activities, such as social media or fitness tracking platforms. These examples also provide an opportunity to highlight that tools for accessing and transforming such data are available in both R and Python, reinforcing the transferability of these skills across languages. While not all educators may have prior experience with these workflows, generative AI tools now offer potential support for developing teaching materials and examples, making it increasingly feasible to incorporate such topics into statistics teaching.

Beyond data structure, the scale of modern datasets can itself pose challenges. Some datasets are simply too large to be handled efficiently on personal machines. While the use of high-performance computing and clusters is common in statistical research, it is less frequently addressed in undergraduate teaching. As datasets continue to grow, distributed computing frameworks become increasingly relevant. Technologies such as Hadoop and Spark can therefore play a role in statistics education where the relevant educator expertise exists. A gradual introduction is often appropriate, beginning with data manipulation and database concepts on a single machine, before progressing to distributed frameworks and cloud-based workflows in more advanced courses.

Computational challenges are not limited to data storage and processing. Many modern ML methods, particularly deep learning models underlying tools such as generative AI systems, require substantial computational resources and are commonly trained using graphical processing units (GPUs). Access to such resources is increasingly mediated through cloud computing platforms, which allow users to submit jobs to remote servers equipped with specialised hardware. Introducing students to cloud-based environments, for example through free platforms such as Google Colab, can provide a practical entry point to these concepts, although the extent to which such material can be incorporated may be constrained by the availability of educators with the appropriate computational expertise. From there, teaching can progress to ideas such as parallelisation, batch processing, and scalable computation. Alongside these technical considerations, it is also important to address ethical issues, including energy consumption, environmental impact, and bias in large-scale modelling.

\subsection{How can we improve the way we teach modern data}
\label{sec:data_teaching}

As outlined above, modern data arise from a wide range of sources and are stored in a variety of formats. While these structures can initially appear complex, many can be read directly into R and processed into forms suitable for statistical analysis. A suite of R packages supports the transformation of modern data formats into standard data frame or tibble representations, expanding the range of data practices that can be incorporated into statistics teaching. The following discussion outlines several common data structures and illustrates how they can be handled within R, while noting that comparable workflows are also available in Python. Together, these tools provide opportunities to extend statistics curricula to better reflect contemporary data acquisition and preparation practices.

One of the most common mechanisms for obtaining modern data is through application programming interfaces (APIs), which allow data to be sent and received via URL-based HTTP requests. In R, API access is supported by packages such as httr2 \citep{httr2}, with endpoints that may be publicly accessible or require authentication through API keys. Data returned from APIs are most often structured in hierarchical formats such as JSON or XML. Packages including jsonlite \citep{jsonlite} and xml2 \citep{xml2} provide straightforward tools for transforming these formats into data frames or tibbles for subsequent analysis. Introducing API-based data access within the curriculum allows students to engage with real-world data sources reinforcing data wrangling skills.

In cases where data are available online but not obtainable through an API, web scraping provides an alternative means of collecting data. Web scraping involves automatically extracting information from web pages and transforming it into structured datasets. In R, the rvest package \citep{rvest} supports the extraction of content from HTML documents by allowing users to identify and select specific elements, commonly referred to as nodes, within a page. Tools such as SelectorGadget from rvest can assist in identifying these elements, making the process more intuitive to learners, avoiding the need to directly interpret HTML code. For websites that rely heavily on dynamically loaded content, particularly by utilising JavaScript, rvest also includes functionality to enable running a live browser session to retrieve such data dynamically. Similar approaches exist in Python using dedicated web scraping and browser automation libraries. Introducing web scraping selectively within statistics programmes can expose students to less structured data sources, while also providing a natural context for discussing ethical, legal, and practical considerations when processing data.

To incorporate modern data techniques effectively within a statistics curriculum, a scaffolded approach is likely to be most appropriate. Rather than treating these topics as stand-alone additions, elements of modern data acquisition and processing can be woven throughout the programme. In the early stages of a degree, structured data remains a natural starting point, supporting the introduction of core statistical concepts alongside basic programming. As students’ computational skills develop, semi-structured data accessed through APIs can be introduced within modelling and programming courses, allowing learners to engage with the full pipeline from data acquisition to analysis. In later years, more complex data sources, including dynamic web content and cloud-based workflows, can be explored within advanced modelling courses, shifting the emphasis from the use of tools to the design of analytical frameworks. This staged approach naturally distributes content across multiple modules, meaning that no single specialist educator can cover all aspects; instead, educators may need to develop familiarity with related topics over time, supported through collaboration and knowledge sharing across teaching teams.

\subsection{Changes in code-based good practice and transparency}
\label{sec:github}

Open-source coding practices are becoming an increasingly important component of software and model development across many areas of statistics. Growing expectations around transparency, code sharing, data availability, and collaborative development have led to greater emphasis on good practice in how code is written, managed, and shared. The sharing and reproducibility of code is now common in many journal publications, for example within the journals of the Royal Statistical Society, and the use of version control systems in research projects continues to expand, with evidence that this can enhance the relevance and longevity of published work \citep{kang2023papers}. Version control platforms such as GitHub, built on the Git system, provide widely adopted mechanisms for tracking changes, collaborating on code, and managing research software. Familiarity with such tools is increasingly expected in industry-facing roles within statistics and data analytics, making exposure to version control an important consideration within statistics education.

The effective integration of version control into the statistics curriculum, however, requires a careful pedagogical design. Version control systems were originally developed for large-scale software engineering projects, and many existing resources assume a background in computer science or software development. For students whose primary focus is statistics or data analysis, learning materials should therefore be tailored to relevant use cases, such as managing individual analysis projects, collaborating on group-based coursework, or contributing to shared research code. An additional consideration is the environment in which version control is introduced. For example, integrating version control through an RStudio-based workflow can reduce cognitive load for beginners by embedding these concepts within a familiar interface, although this may limit the applicability of those skills to a narrower range of future projects. Alternatively, introducing version control directly through platforms such as GitHub Desktop may provide more broadly transferable skills, but can present a steeper initial learning curve. Decisions about which approach to adopt, and when, should therefore be informed by students’ prior experience and the intended learning outcomes of the programme.

Once introduced, incorporating version control within assessment can support skill development by giving students opportunities for authentic practice. For example, group-based data analysis projects can use version control to support collaborative coding, shared analysis, and peer feedback throughout the project lifecycle, including code review via GitHub pull requests. Beyond technical skills, such approaches can also enable more robust assessment design in the context of generative AI, as discussed further in Section~\ref{sec:genAI}. In online programming courses, for example, where access to generative AI tools cannot be fully restricted, assessment of core concepts can be supported through interaction with non-public code libraries hosted on platforms such as GitHub, requiring students to engage directly with existing code and extend its functionality. Peer review activities can further reinforce practical experience with version control, code review, and collaborative development workflows while supporting academic integrity.


\section{Incorporating ML and AI into Statistical Education}
\label{sec:ML_AI}

The distinction between statistics, ML, and AI is often unclear \citep{ij18}, with substantial overlap between the respective fields. High-profile textbooks frequently cover material that spans multiple disciplines \citep{bishop06,murphy12}. For example, methods such as the LASSO and elastic net are often regarded as both statistical and ML techniques. Similarly, deep learning, typically classified under AI, incorporates statistical regression models as part of its core structure, yet is rarely considered part of statistics despite well-recognised connections \citep{cheng94,white89}. While we do not seek to impose rigid disciplinary boundaries, working definitions are necessary to provide context for discussion. In this article, we define AI as methods based on neural networks, and ML as modelling approaches not traditionally included in statistics curricula, excluding neural networks. We also avoid the common, but arguably misleading, distinction of statistics as causal and ML or AI as predictive, as many modern methods blur this divide \citep{scholkopf22}. These definitions are intentionally flexible, providing a coherent framework for discussion while allowing educators to adapt them to their specific curricula and student needs.

From an educational perspective, given the overlap between these fields, it can be argued that attempting to rigidly define boundaries between statistics, ML, and AI is neither necessary nor particularly helpful. An alternative approach is to present multiple perspectives, allowing students to develop a more nuanced understanding of how these areas relate to one another. In this view, ML and AI need not be positioned as competitors to statistics, but rather as increasingly integral components of many statistical roles. When included in the curriculum, they may be taught with a balanced emphasis on both strengths and limitations. The extent and depth of their inclusion will depend on a programme’s focus, intended learning outcomes, and desired graduate attributes. For example, programmes aimed at clinical statistics might include limited coverage of ML and AI, with greater emphasis placed on their constraints and appropriate use cases \citep{wilkinson20}. In contrast, data science-oriented programmes are likely to require more comprehensive coverage, alongside strong statistical foundations to help students avoid common methodological pitfalls \citep{arnold20}. Ultimately, the extent to which such content can be incorporated depends on the availability of educators with the relevant expertise to teach it effectively.

\subsection{Teaching ML}
\label{sec:ML}

A central aim of higher education is to prepare students for further study or employment. From this perspective, decisions about whether to include ML in a statistics curriculum should be guided by likely graduate pathways. For students pursuing careers involving substantial statistical work, such as academia, data science, or applied analytics, some exposure to ML is likely to be important, as these methods increasingly form part of routine practice. By contrast, for students in disciplines outside the mathematical sciences, such as psychology or medicine, regular engagement with ML may be less common. In such cases, it may be reasonable to limit formal training in ML. However, one could still argue for educating these students about the potential risks of poorly applied ML, particularly as the field grows in popularity and diverse applications, enabling them to critically interpret research findings and avoid common methodological pitfalls within their fields \citep{wilkinson20}.

The widespread use of ML methods across applied disciplines, sometimes without sufficient methodological rigour, further supports the case for at least a conceptual introduction to their strengths and limitations \citep{wilkinson20}. Many of these issues align closely with core statistical principles, including non-linear effects, overfitting, and model validation. From this standpoint, introducing ML provides an opportunity to reinforce existing statistical concepts, or at minimum to raise awareness of where and why problems may arise. More broadly, well-designed ML analyses depend heavily on statistical reasoning \citep{arnold20}, including considerations of causality, uncertainty, and critical evaluation. Addressing these ideas explicitly can therefore offer a pragmatic way to situate ML within a statistics-led curriculum.

If ML is to be incorporated into a programme, careful thought is required as to how this should be done. Unlike the parallel teaching of R and Python discussed in Section~\ref{sec:Combine_R_Python}, statistics and ML should probably not be considered interchangeable, despite sometimes being treated that way in practice. As a result, approaches such as the language switcher illustrated in Figure~\ref{fig:switcher} are not directly applicable. One option is to integrate ML topics within existing courses where there is a natural conceptual overlap, for example introducing Gaussian processes within a non-linear modelling course traditionally focused on teaching generalised additive models. Alternatively, a standalone ML course may be appropriate, provided it is clearly grounded in statistical concepts already covered elsewhere in the curriculum. Taken together, a combination of targeted integration and dedicated teaching has the potential to offer a balanced and flexible approach. This may lead to course titles that challenge disciplinary distinctions, such as ‘ML Methods’, but such compromises reflect the broader challenges associated with defining statistics, ML, and AI, and can also provide opportunities to open discussion with students about how these fields relate to one another.

\subsection{Teaching AI}
\label{sec:AI}

When considering ML education, a natural question is whether AI methods should also be included. Deep learning approaches, which build on multi-layer neural networks, are increasingly prevalent in both academic research and industry, suggesting that some level of exposure may be beneficial for students. This raises further questions about how AI should be positioned within statistics curricula and how it relates to existing ML content.

In practice, the teaching of AI will often be constrained by staff expertise, although this may change as AI methods become more embedded within statistical research. At an introductory level, students can be exposed to deep learning methods for tasks such as classification and regression without engaging in detailed theoretical treatments of underlying mechanisms such as backpropagation or automatic differentiation. Demonstrating the practical use of these methods in R or Python could require only a high-level understanding of the models and can be incorporated within existing ML courses.

More advanced coverage of AI is possible, but presents additional challenges in terms of both staff expertise and students’ technical backgrounds. Topics such as large language models and transformer architectures, while increasingly prominent, would be best approached at an applied level if they were to be included in statistics programmes. Similarly, more specialised skills, including the use of Linux-based systems, graphical processing units (GPUs), and parallel computing, might typically more appropriate as advanced or supplementary topics, or as part of specialist training outside the core statistics curriculum. In cases where deeper technical expertise in AI is required, it may be more effective for such training to be delivered in collaboration with, or delegated to, related disciplines such as computer science or engineering.


\section{The emergence of Generative AI tools}
\label{sec:genAI}

Generative AI tools such as ChatGPT make rapidly advancing large language models readily accessible through interactive interfaces, lowering the barrier to their use in educational and professional contexts. As a result, changes are already evident in how students approach learning new topics and engage with assessment, alongside indications that patterns of student engagement are shifting in sometimes unpredictable ways \citep{pardos2023learning}. In response, it is increasingly apparent that learning and teaching practice must adapt to these developments and, in some capacity, engage constructively with such tools.

A central question then becomes how, and to what extent, generative AI should be addressed within statistics education. One option is to focus primarily on the risks associated with these tools, discouraging or restricting their use. An alternative approach is to acknowledge their widespread adoption and encourage students to use them responsibly, while explicitly teaching their limitations, potential biases, and failure modes. This latter approach raises further questions about what constitutes responsible use, how such practices should be incorporated into teaching and assessment, and how educators can develop the expertise needed to support students effectively in this area.

From the perspective of graduate attributes, the growing reliance on generative AI tools within industry suggests that some level of engagement is increasingly difficult to avoid. Evidence also indicates that students are already using such tools for a wide range of tasks \citep{freeman2025student}, although there is also some suggestion that educators may overestimate the extent of student usage \citep{lee2024impact}. In this context, it can be argued that developing guidance on how to teach effective generative AI usage is more constructive than attempting to prohibit its use outright. This remains an active area of discussion within the literature, with several proposed frameworks and guiding principles for the responsible integration of generative AI into higher education \citep{nartey2024guiding,grove2025designing}.

From a learning and teaching perspective, the adoption of generative AI presents a number of challenges. Assessment design is a particularly prominent concern, especially as students rapidly adopt these technologies across academia. Equally important is how students are supported in developing effective and critical approaches to using generative AI. Teaching staff also require support in identifying potential misuse, understanding how these tools can inform their own practice, and developing confidence in guiding students towards responsible use. The following sections explore specific challenges posed by generative AI tools and consider their implications for statistics education in greater depth.

\subsection{Teaching about Generative AI tools}
\label{sec:teach_genai}

Teaching around generative AI presents a number of challenges, not least because the technology itself is evolving rapidly. Until relatively recently, effective use of generative AI tools often required a detailed understanding of prompt engineering, that is, the process by which instructions are formulated to elicit useful responses from a generative model. Increasingly, however, these tools are able to generate plausible outputs in response to less precise or carefully structured queries \citep{wang2024advanced}. From an educational perspective, this ever changing landscape creates difficulties in maintaining up-to-date teaching materials, particularly within institutional contexts where educators manage a range of competing responsibilities.

Despite these tools becoming easier to use, it remains clear that there are still both effective and ineffective ways of engaging with them. Students therefore need to be supported in developing an understanding of how generative AI tools function, including what they can and cannot do reliably. In a statistics context, this may be most effectively achieved by embedding discussion of generative AI within statistical or ML analyses, recognising that such tools may be effective for certain components of the workflow while less appropriate for others. For example, tools may be well suited to supporting coding tasks or producing visualisations, while being inappropriate for more substantive decisions such as covariate selection or model choice. Determining how best to convey these distinctions, and how much emphasis to place on them, represents an ongoing challenge across the sector.

Beyond technical use, the adoption of generative AI raises broader considerations that are also relevant to statistics education. Universities are increasingly developing formal guidance on how such tools may be used and how their use should be declared. While it is necessary for educators to communicate institutional policies, there is also a case for addressing these issues at a more general level. Students who are not encouraged to reflect on how and why they use generative AI tools are unlikely to consider the wider implications of their use. These broader considerations include the environmental impact of generative AI systems \citep{rillig2023risks}, ethical issues such as transparency, bias, and accountability \citep{Hagendorff_2024}, and ongoing concerns around the use of copyrighted material in training data and generated outputs \citep{buick2025copyright}, all of which universities may reasonably be expected to address within higher education.

Taken together, these considerations raise the question of where and how education about generative AI should be situated within a statistics programme. A stand-alone course focused solely on generative AI, distinct from foundational AI or ML content, more extensive than is currently necessary. At the same time, ignoring these tools entirely is no longer seems a viable option. Embedding discussion of generative AI throughout all courses, beyond communicating institutional rules, may also place unreasonable demands on both students and staff. A more pragmatic approach may be to situate education about generative AI within skills-based modules, in a manner comparable to how communication, presentation, and consultancy skills are typically addressed within statistics education.

\subsection{Student usage of Generative AI tools}
\label{sec:student_usage_genai}

The extent to which students should use generative AI tools, and whether such use ultimately benefits their education, remains an open question. A growing body of literature has begun to explore this issue in greater depth than is possible here, e.g., \cite{lehmann2024ai}, but existing studies report a wide range of outcomes and vary substantially in scope, methodology, and research design. Even where positive effects have been observed, it remains unclear how far these findings can be generalised. In particular, it is difficult to assess the long-term implications of students completing an entire degree programme in sustained conjunction with generative AI tools.

From an educational perspective, this uncertainty makes it challenging to determine where and how the use of generative AI is pedagogically appropriate. These decisions must be made not only at the programme level but also within individual modules and learning activities. In statistics education, for example, there are emerging case studies that report promising outcomes from the use of generative AI tools \citep{al2025enhancing}. At the same time, there is concern that sustained reliance on such tools, particularly as their use expands across multiple courses, may contribute to a form of apathy in which students become less engaged in the effortful reasoning needed to develop statistical intuition, potentially weakening long-term conceptual understanding \citep{AIapathy}. However, comprehensive comparisons across a broad range of implementations, course formats, and learning contexts are not yet available, and the cumulative impact of repeated exposure across a degree programme remains largely unstudied, reflecting the fact that generative AI tools have only recently become widely adopted.

A further consideration is whether generative AI agents should be integrated directly into educational platforms. Such integration could represent a substantial shift in how education is delivered, offering the possibility of highly accessible, responsive AI-based tutors that help bridge gaps in understanding and provide more immediate support than is typically feasible for academic staff. At the same time, closer integration raises concerns about the propagation of incorrect information and the risk of increased over-reliance on automated systems. Given that many students are already using generative AI tools informally, it remains unclear whether tighter institutional integration represents a natural progression or a step towards even greater over-reliance on these technologies.

\subsection{Assessment, marking, and feedback}
\label{sec:genai_assessment}

The rapid rise of generative AI tools has created significant challenges for assessment across much of higher education. Evidence suggests that many existing assessment formats are vulnerable and may require relatively rapid adaptation \citep{newton2025vulnerable}, with certain types of assessment appearing particularly exposed to misuse \citep{grove2024generative}. A commonly proposed response is a return to invigilated, on-campus examinations. However, such approaches are not universally welcomed within higher education and have well-documented strengths and limitations in terms of assessing student learning \citep{buckley2024we}. As a result, there has been growing interest in using the emergence of generative AI as an opportunity to reconsider broader principles of assessment design and to establish new standards of good practice \citep{grove2024generative}, although it remains unclear what such assessments should look like in practice.

Interestingly, generative AI tools may also present opportunities within assessment, particularly from the perspective of efficiency and consistency. Some educators have begun to explore the use of generative AI in the creation of assessment materials, with emerging guidance on how this might be done effectively in the context of statistics education \citep{gordon2026ai}. While this offers clear advantages in terms of efficiency of time management, it also raises concerns around accuracy, including the potential for hallucinated content or insufficient alignment with course-specific context. More advanced possibilities include the use of fine-tuned models trained on course materials, which may offer greater contextual awareness. In addition, while caution is clearly required when using generative AI as a content-generation tool, there is also potential value in employing these systems as supplementary checks on assessment materials, for example to identify obvious errors or ambiguities prior to release.

Opportunities also exist in the areas of marking and feedback. A number of frameworks have been proposed to support the use of generative AI as a marking aid \citep{safilian2025ratas}, including work focused specifically on statistics education \citep{ilieva2025framework}. Given the quantitative and computational nature of the discipline, there may be particular scope for supporting the assessment of code-based work. Generative AI tools may also enable streamlining processes to provide more personalised feedback in contexts where this would otherwise be impractical, such as in large classes. At the same time, concerns remain around trust, transparency, and perceived fairness. If students or staff lose confidence in the reliability of AI-supported marking and feedback, there is a risk of undermining the educational process, particularly in settings where the financial and personal costs of higher education are substantial.


\section{Summary and Discussion}
\label{sec:future_challenges}

This article has examined recent technological developments and considered their implications for contemporary statistics education. In particular, it has explored how changes in programming practice have shaped teaching, from the emergence of R and its transformative impact on statistical computing to the growing need to consider how Python may be integrated into certain statistics curricula, particularly those with a focus on data science or ML. Approaches for teaching R and Python in parallel have also been discussed as a pragmatic response to modern analytical practice.

The paper has further considered the influence of modern data sources and coding practices on statistics education. As data types become more diverse and complex, there is a growing need for curricula to adapt accordingly. Rather than isolating these topics within standalone modules, it has been argued that modern data handling practices should, where possible, be integrated throughout degree programmes. Related to this is the increasing importance of version control, driven both by industry expectations and the need to develop appropriate graduate attributes within statistics education.

The integration of ML and AI into statistics curricula has also been examined. Given the stated aim of higher education to prepare students for future academic and professional pathways, it has been argued that decisions about inclusion should be guided by the likely destinations of graduates. A foundational understanding of ML, particularly its strengths and limitations, appears increasingly important across many statistics programmes. Broader or more advanced coverage of AI, defined here as methods based on neural networks, should be contingent on the availability of appropriate expertise and their relevance to future careers or further study. In cases where more specialised AI training is required, this may be better delivered by related disciplines or considered beyond the scope of a statistics programme.

Finally, the paper has discussed the growing impact of generative AI tools, such as ChatGPT, on statistical learning and assessment. These tools are already influencing how students engage with learning and assessment, and this influence is likely to continue irrespective of whether institutions actively embrace them. From the perspective of graduate attributes, and given evidence that many students are already using generative AI tools \citep{freeman2025student}, there is a strong case for providing structured education about their use. This includes not only discussion of risks and limitations, but also guidance on responsible and effective use. How best to achieve this remains an open question, although situating such content within skills-focused modules has been proposed as a pragmatic approach. Wider implications for assessment and student learning have also been highlighted.

Taken together, these discussions underline that statistics education has a long history of adapting to methodological and technological change. Educators now face the challenge of responding to rapid developments in analytical methods, including ML and AI, evolving data types such as text, images, and web-based sources, and changing expectations around coding skills, including Python, version control, and cloud-based computation. Determining how these elements should be incorporated will necessarily depend on the specific curriculum, student cohort, and programme aims. In this context, a diversity of skills across teaching teams becomes increasingly valuable, helping to ensure that expertise is distributed rather than concentrated in a small number of individuals. Supporting educators in finding time and opportunities to upskill is therefore essential. Research activity can play an important role in this process, but professional development, collaboration, and shared teaching practices are equally important for sustaining long-term curriculum development.

In conclusion, statistics education is undergoing significant change, bringing challenges for students, educators, and institutions alike. Keeping pace with evolving technologies and methods requires not only thoughtful curriculum design but also continued reflection on how statistical thinking, computational skills, and emerging tools can coexist within coherent educational frameworks. Rather than representing a departure from the discipline’s foundations, these developments highlight the ongoing need for adaptable, critically informed teaching that prepares students for a rapidly changing analytical landscape.

\section*{Orcids}

Craig Alexander - \url{https://orcid.org/0000-0001-6734-747X} \\
Jennifer Gaskell - \url{https://orcid.org/0000-0001-9583-323X} \\
Vinny Davies - \url{https://orcid.org/0000-0003-1896-8936}

\bibliographystyle{abbrvnat}
\bibliography{main}

@article{AIapathy,
	author = {Fan, Lei and Deng, Kunyang and Liu, Fangxue},
	doi = {10.1038/s41598-025-06930-w},
	journal = {Scientific Reports},
	number = {1},
	pages = {26521},
	title = {Educational impacts of generative artificial intelligence on learning and performance of engineering students in {C}hina},
	volume = {15},
	year = {2025},
}

@article{white89,
  title={Learning in artificial neural networks: A statistical perspective},
  author={White, Halbert},
  journal={Neural computation},
  volume={1},
  number={4},
  pages={425--464},
  year={1989},
  publisher={MIT Press One Rogers Street, Cambridge, MA 02142-1209, USA journals-info~…}
}

@article{cheng94,
  title={Neural networks: A review from a statistical perspective},
  author={Cheng, Bing and Titterington, D Michael},
  journal={Statistical science},
  pages={2--30},
  year={1994},
  publisher={JSTOR}
}

@article{ij18,
  title={Statistics versus machine learning},
  author={Bzdok, Danilo and Altman, Naomi and Krzywinski, Martin},
  journal={Nat Methods},
  volume={15},
  number={4},
  pages={233},
  year={2018},
  doi={10.1038/nmeth.4642}
}

@book{bishop06,
  title={Pattern recognition and machine learning},
  author={Bishop, Christopher M},
  volume={4},
  number={4},
  year={2006},
  publisher={Springer}
}

@book{murphy12,
  title={Machine learning: a probabilistic perspective},
  author={Murphy, Kevin P},
  year={2012},
  publisher={MIT press}
}

@article{arnold20,
  title={Reflection on modern methods: generalized linear models for prognosis and intervention—theory, practice and implications for machine learning},
  author={Arnold, Kellyn F and Davies, Vinny and de Kamps, Marc and Tennant, Peter WG and Mbotwa, John and Gilthorpe, Mark S},
  journal={International journal of epidemiology},
  volume={49},
  number={6},
  pages={2074--2082},
  year={2020},
  publisher={Oxford University Press}
}

@book{scholkopf22,
   title={Causality for Machine Learning},
   ISBN={9781450395861},
   url={http://dx.doi.org/10.1145/3501714.3501755},
   DOI={10.1145/3501714.3501755},
   booktitle={Probabilistic and Causal Inference},
   publisher={ACM},
   author={Schölkopf, Bernhard},
   year={2022},
   month=feb, pages={765–804} }

@Article{tidyverse19,
    title = {Welcome to the {tidyverse}},
    author = {Hadley Wickham and Mara Averick and Jennifer Bryan and Winston Chang and Lucy D'Agostino McGowan and Romain François and Garrett Grolemund and Alex Hayes and Lionel Henry and Jim Hester and Max Kuhn and Thomas Lin Pedersen and Evan Miller and Stephan Milton Bache and Kirill Müller and Jeroen Ooms and David Robinson and Dana Paige Seidel and Vitalie Spinu and Kohske Takahashi and Davis Vaughan and Claus Wilke and Kara Woo and Hiroaki Yutani},
    year = {2019},
    journal = {Journal of Open Source Software},
    volume = {4},
    number = {43},
    pages = {1686},
    doi = {10.21105/joss.01686},
  }

@article{wickham14,
  title={Tidy data},
  author={Wickham, Hadley},
  journal={Journal of statistical software},
  volume={59},
  pages={1--23},
  year={2014},
  doi={10.18637/jss.v059.i10}
}

@article{wilkinson20,
  title={Time to reality check the promises of machine learning-powered precision medicine},
  author={Wilkinson, Jack and Arnold, Kellyn F and Murray, Eleanor J and van Smeden, Maarten and Carr, Kareem and Sippy, Rachel and de Kamps, Marc and Beam, Andrew and Konigorski, Stefan and Lippert, Christoph and others},
  journal={The Lancet Digital Health},
  volume={2},
  number={12},
  pages={e677--e680},
  year={2020},
  publisher={Elsevier}
}

@article{Kalyan18,
    title = {Python vs. {R} Programming Language},
    author = {Kalyan Sudhaka},
    journal = {International Journal of Management, IT and Engineering},
    volume = {8},
    number = {8},
    pages = {},
    year = {2018}
    }

@article{ozgur2017b,
  title={MatLab vs. {P}ython vs. {R}},
  author={Ozgur, Ceyhun and Colliau, Taylor and Rogers, Grace and Hughes, Zachariah and others},
  journal={Journal of data Science},
  volume={15},
  number={3},
  pages={355--371},
  year={2017}
}

@book{wickham23,
  title = {R for Data Science: Import, Tidy, Transform, Visualize, and Model Data},
  author = {Hadley Wickham and Mine Çetinkaya-Rundel and Garrett Grolemund},
  year = {2023},
  publisher = {O'Reilly Media},
  edition = {2nd},
  url = {https://r4ds.hadley.nz}
}

@book{ryan21,
  title={Data Science with R for Psychologists and Healthcare Professionals},
  author={Ryan, Christian},
  year={2021},
  publisher={CRC Press}
}

@book{imai22,
  title={Quantitative Social Science: An Introduction in Tidyverse},
  author={Imai, Kosuke and Williams, Nora Webb},
  year={2022},
  publisher={Princeton University Press}
}

@book{musa23,
  title={Data Analysis in Medicine and Health Using R},
  author={Musa, Kamarul Imran and Mansor, Wan Nor Arifin Wan and Hanis, Tengku Muhammad},
  year={2023},
  publisher={CRC Press}
}

@Manual{shiny,
    title = {shiny: Web Application Framework for R},
    author = {Winston Chang and Joe Cheng and JJ Allaire and Carson Sievert and Barret Schloerke and Yihui Xie and Jeff Allen and Jonathan McPherson and Alan Dipert and Barbara Borges},
    year = {2025},
    note = {R package version 1.11.1},
    url = {https://CRAN.R-project.org/package=shiny},
    doi = {10.32614/CRAN.package.shiny},
  }

@article{potter16,
  title={Web application teaching tools for statistics using R and shiny},
  author={Potter, Gail and Wong, Jimmy and Alcaraz, Irvin and Chi, Peter and others},
  journal={Technology Innovations in Statistics Education},
  volume={9},
  number={1},
  year={2016}
}

@article{fawcett18,
  title={Using interactive shiny applications to facilitate research-informed learning and teaching},
  author={Fawcett, Lee},
  journal={Journal of Statistics Education},
  volume={26},
  number={1},
  pages={2--16},
  year={2018},
  publisher={Taylor \& Francis}
}

@Manual{rmarkdown,
  title = {rmarkdown: Dynamic Documents for R},
  author = {JJ Allaire and Yihui Xie and Christophe Dervieux and Jonathan McPherson and Javier Luraschi and Kevin Ushey and Aron Atkins and Hadley Wickham and Joe Cheng and Winston Chang and Richard Iannone},
  year = {2024},
  note = {R package version 2.29},
  url = {https://github.com/rstudio/rmarkdown},
}

@Manual{quarto,
    title = {quarto: R Interface to `Quarto' Markdown Publishing System},
    author = {JJ Allaire and Christophe Dervieux},
    year = {2025},
    note = {R package version 1.5.1},
    url = {https://CRAN.R-project.org/package=quarto},
    doi = {10.32614/CRAN.package.quarto},
  }

@software{pandas,
    author       = {Pandas},
    title        = {pandas-dev/pandas: Pandas},
    month        = feb,
    year         = 2020,
    publisher    = {Zenodo},
    version      = {latest},
    doi          = {10.5281/zenodo.3509134},
    url          = {https://doi.org/10.5281/zenodo.3509134}
}

@Article{matplotlib,
  Author    = {Hunter, J. D.},
  Title     = {Matplotlib: A 2D graphics environment},
  Journal   = {Computing in Science \& Engineering},
  Volume    = {9},
  Number    = {3},
  Pages     = {90--95},
  publisher = {IEEE COMPUTER SOC},
  doi       = {10.1109/MCSE.2007.55},
  year      = 2007
}

@Book{ggplot2,
    author = {Hadley Wickham},
    title = {ggplot2:Elegant Graphics for Data Analysis},
    publisher = {Springer-Verlag New York},
    year = {2016},
    isbn = {978-3-319-24277-4},
    url = {https://ggplot2.tidyverse.org},
  }

@Book{mgcv,
    title = {Generalized Additive Models: An Introduction with R},
    year = {2017},
    author = {S.N Wood},
    edition = {2},
    publisher = {Chapman and Hall/CRC},
  }

@misc{tensorflow,
title={{TensorFlow}: Large-Scale Machine Learning on Heterogeneous Systems},
url={http://tensorflow.org/},
note={Software available from tensorflow.org},
author={
    Mart\'{\i}n~Abadi and
    Ashish~Agarwal and
    Paul~Barham and
    Eugene~Brevdo and
    Zhifeng~Chen and
    Craig~Citro and
    Greg~S.~Corrado and
    Andy~Davis and
    Jeffrey~Dean and
    Matthieu~Devin and
    Sanjay~Ghemawat and
    Ian~Goodfellow and
    Andrew~Harp and
    Geoffrey~Irving and
    Michael~Isard and
    Yangqing Jia and
    Rafal~Jozefowicz and
    Lukasz~Kaiser and
    Manjunath~Kudlur and
    Josh~Levenberg and
    Dan~Man\'{e} and
    Rajat~Monga and
    Sherry~Moore and
    Derek~Murray and
    Chris~Olah and
    Mike~Schuster and
    Jonathon~Shlens and
    Benoit~Steiner and
    Ilya~Sutskever and
    Kunal~Talwar and
    Paul~Tucker and
    Vincent~Vanhoucke and
    Vijay~Vasudevan and
    Fernanda~Vi\'{e}gas and
    Oriol~Vinyals and
    Pete~Warden and
    Martin~Wattenberg and
    Martin~Wicke and
    Yuan~Yu and
    Xiaoqiang~Zheng},
  year={2015},
}

@software{jax,
  author = {James Bradbury and Roy Frostig and Peter Hawkins and Matthew James Johnson and Chris Leary and Dougal Maclaurin and George Necula and Adam Paszke and Jake Vander{P}las and Skye Wanderman-{M}ilne and Qiao Zhang},
  title = {{JAX}: composable transformations of {P}ython+{N}um{P}y programs},
  url = {http://github.com/jax-ml/jax},
  version = {0.3.13},
  year = {2018},
}

@inproceedings{pytorch,
 author = {Paszke, Adam and Gross, Sam and Massa, Francisco and Lerer, Adam and Bradbury, James and Chanan, Gregory and Killeen, Trevor and Lin, Zeming and Gimelshein, Natalia and Antiga, Luca and Desmaison, Alban and Kopf, Andreas and Yang, Edward and DeVito, Zachary and Raison, Martin and Tejani, Alykhan and Chilamkurthy, Sasank and Steiner, Benoit and Fang, Lu and Bai, Junjie and Chintala, Soumith},
 booktitle = {Advances in Neural Information Processing Systems},
 publisher = {Curran Associates, Inc.},
 title = {PyTorch: An Imperative Style, High-Performance Deep Learning Library},
 url = {https://proceedings.neurips.cc/paper_files/paper/2019/file/bdbca288fee7f92f2bfa9f7012727740-Paper.pdf},
 volume = {32},
 year = {2019}
}

@Manual{httr2,
    title = {httr2: Perform HTTP Requests and Process the Responses},
    author = {Hadley Wickham},
    year = {2023},
    note = {R package version 0.2.3},
    url = {https://CRAN.R-project.org/package=httr2},
  }

@Article{jsonlite,
    title = {The jsonlite {P}ackage: {A} {P}ractical and {C}onsistent {M}apping {B}etween {JSON} {D}ata and {R} {O}bjects},
    author = {Jeroen Ooms},
    journal = {arXiv:1403.2805 [stat.CO]},
    year = {2014},
    url = {https://arxiv.org/abs/1403.2805},
  }

@Manual{xml2,
    title = {xml2: Parse XML},
    author = {Hadley Wickham and Jim Hester and Jeroen Ooms},
    year = {2023},
    note = {R package version 1.3.6},
    url = {https://CRAN.R-project.org/package=xml2},
  }

@Manual{rvest,
    title = {rvest: Easily Harvest (Scrape) Web Pages},
    author = {Hadley Wickham},
    year = {2022},
    note = {R package version 1.0.3},
    url = {https://CRAN.R-project.org/package=rvest},
  }

@article{diggle2015statistics,
  title={Statistics: a data science for the 21st century},
  author={Diggle, Peter J},
  journal={Journal of the Royal Statistical Society Series A: Statistics in Society},
  volume={178},
  number={4},
  pages={793--813},
  year={2015},
  publisher={Oxford University Press}
}

@article{hardin2015data,
  title={Data science in statistics curricula: Preparing students to “think with data”},
  author={Hardin, Johanna and Hoerl, Roger and Horton, Nicholas J and Nolan, Deborah and Baumer, Ben and Hall-Holt, Olaf and Murrell, Paul and Peng, Roger and Roback, Paul and Temple Lang, D and others},
  journal={The American Statistician},
  volume={69},
  number={4},
  pages={343--353},
  year={2015},
  publisher={Taylor \& Francis}
}

@book{kenett2022modern,
  title={Modern statistics: a computer-based approach with python},
  author={Kenett, Ron S and Zacks, Shelemyahu and Gedeck, Peter},
  year={2022},
  publisher={Springer}
}

@article{bond20,
  title={Mapping research in student engagement and educational technology in higher education: A systematic evidence map},
  author={Bond, Melissa and Buntins, Katja and Bedenlier, Svenja and Zawacki-Richter, Olaf and Kerres, Michael},
  journal={International journal of educational technology in higher education},
  volume={17},
  pages={1--30},
  year={2020},
  publisher={Springer}
}

@article{granic22,
  title={Educational technology adoption: A systematic review},
  author={Grani{\'c}, Andrina},
  journal={Education and Information Technologies},
  volume={27},
  number={7},
  pages={9725--9744},
  year={2022},
  publisher={Springer}
}

@article{johnson97,
  title={Teaching {C}reative {P}roblem {S}olving and {A}pplied {R}easoning {S}kills: {A} {M}odular {A}pproach},
  author={Johnson, Andrea L},
  journal={Cal. WL Rev.},
  volume={34},
  pages={389},
  year={1997},
  publisher={HeinOnline}
}

@article{wickham11,
author = {Wickham, Hadley},
title = {ggplot2},
journal = {WIREs Computational Statistics},
volume = {3},
number = {2},
pages = {180-185},
doi = {https://doi.org/10.1002/wics.147},
year = {2011}
}

@software{plotnine,
  author = {Plotnine},
  title  = {plotnine: A grammar of graphics for Python},
  url    = {https://github.com/has2k1/plotnine},
  doi    = {https://doi.org/10.5281/zenodo.1325308},
  year = {2026}
}

@misc{pyGAM, 
    title={pyGAM: Generalized Additive Models in Python}, 
    DOI={10.5281/zenodo.1208724}, 
    publisher={Zenodo}, 
    author={Servén, Daniel and Brummitt, Charlie}, 
    year={2018} }

@software{pymgcv,
 title = {pymgcv: Generalized Additive Models in Python},
 author = {Pymgcv},
 url = {https://smoothforge.github.io/pymgcv/},
 year = {2026}
}

@article{jack2023reflections,
  title={Reflections on designing and delivering an online distance learning programme in the mathematical sciences},
  author={Jack, Eilidh and Alexander, Craig and McArthur, David and Mair, Colette},
  journal={MSOR Connections},
  volume={21},
  number={2},
  pages={25--33},
  year={2023},
  doi={10.21100/msor.v21i2.1397}
}

@article{kang2023papers,
  title={Papers with code or without code? Impact of GitHub repository usability on the diffusion of machine learning research},
  author={Kang, Donghyun and Kang, TaeYoung and Jang, Junkyu},
  journal={Information Processing \& Management},
  volume={60},
  number={6},
  pages={103477},
  year={2023},
  publisher={Elsevier}
}

@book{freeman2025student,
  title={Student generative {AI} survey 2025},
  author={Freeman, Josh},
  publisher={Higher Education Policy Institute: London, UK},
  year={2025}
}

@article{nartey2024guiding,
  title={Guiding principles of generative AI for employability and learning in UK universities},
  author={Nartey, Emmanuel K},
  journal={Cogent education},
  volume={11},
  number={1},
  pages={2357898},
  year={2024},
  publisher={Taylor \& Francis}
}

@article{grove2025designing,
  title={Designing the {S}tudent {L}earning {J}ourney: {A} {P}ractical {A}pproach to {I}ntegrating {G}enerative {AI} within {H}igher {E}ducation},
  author={Grove, Michael},
  journal={MSOR Connections},
  volume={24},
  number={1},
  year={2025}
}

@article{lee2024impact,
  title={The impact of generative AI on higher education learning and teaching: A study of educators’ perspectives},
  author={Lee, Daniel and Arnold, Matthew and Srivastava, Amit and Plastow, Katrina and Strelan, Peter and Ploeckl, Florian and Lekkas, Dimitra and Palmer, Edward},
  journal={Computers and Education: Artificial Intelligence},
  volume={6},
  pages={100221},
  year={2024},
  publisher={Elsevier}
}

@article{grove2024generative,
  title={Generative AI technologies and their role within assessment design},
  author={Grove, Michael},
  journal={Education in Practice},
  volume={5},
  number={1},
  pages={15--35},
  year={2024},
  publisher={University of Birmingham}
}

@article{buckley2024we,
  title={Are we answering the question that has been set? Exploring the gap between research and practice around examinations in higher education},
  author={Buckley, Alex},
  journal={Studies in Higher Education},
  volume={49},
  number={11},
  pages={1928--1944},
  year={2024},
  publisher={Taylor \& Francis}
}

@article{wang2024advanced,
  title={Do advanced language models eliminate the need for prompt engineering in software engineering?},
  author={Wang, Guoqing and Sun, Zeyu and Ye, Sixiang and Gong, Zhihao and Chen, Yizhou and Zhao, Yifan and Liang, Qingyuan and Hao, Dan},
  journal={ACM Transactions on Software Engineering and Methodology},
  year={2024},
  publisher={ACM New York, NY}
}

@article{buick2025copyright,
  title={Copyright and {AI} training data—transparency to the rescue?},
  author={Buick, Adam},
  journal={Journal of Intellectual Property Law and Practice},
  volume={20},
  number={3},
  pages={182--192},
  year={2025},
  publisher={Oxford University Press UK}
}

@article{rillig2023risks,
  title={Risks and benefits of large language models for the environment},
  author={Rillig, Matthias C and {\AA}gerstrand, Marlene and Bi, Mohan and Gould, Kenneth A and Sauerland, Uli},
  journal={Environmental science \& technology},
  volume={57},
  number={9},
  pages={3464--3466},
  year={2023},
  publisher={ACS Publications}
}

@article{Hagendorff_2024,
   title={Mapping the Ethics of Generative AI: A Comprehensive Scoping Review},
   volume={34},
   ISSN={1572-8641},
   url={http://dx.doi.org/10.1007/s11023-024-09694-w},
   DOI={10.1007/s11023-024-09694-w},
   number={4},
   journal={Minds and Machines},
   publisher={Springer Science and Business Media LLC},
   author={Hagendorff, Thilo},
   year={2024},
   month=sep }

@article{pardos2023learning,
  title={Learning gain differences between ChatGPT and human tutor generated algebra hints},
  author={Pardos, Zachary A and Bhandari, Shreya},
  journal={arXiv preprint arXiv:2302.06871},
  year={2023}
}

@article{lehmann2024ai,
  title={AI meets the classroom: When does ChatGPT harm learning},
  author={Lehmann, Matthias and Cornelius, Philipp B and Sting, Fabian J},
  journal={arXiv preprint arXiv:2409.09047},
  year={2024}
}

@article{al2025enhancing,
author = {Al Labadi, Luai and Ly, Anna},
title = {Enhancing statistics education through {Project-Based Learning (PBL)} and the emergence of {ChatGPT}},
journal = {Teaching Statistics},
volume = {47},
number = {3},
pages = {200-218},
doi = {https://doi.org/10.1111/test.12405},
year = {2025}
}

@article{newton2025vulnerable,
  title={How vulnerable are UK universities to cheating with new GenAI tools? A pragmatic risk assessment},
  author={Newton, Philip M},
  journal={Assessment \& Evaluation in Higher Education},
  pages={1--12},
  year={2025},
  publisher={Taylor \& Francis}
}

@article{gordon2026ai,
author = {Gordon, Ellen and Chimeli, Janna and Kenyo, Lauren and Frost, Raymond},
title = {From AI to TA: How to use ChatGPT to quickly create statistics and analytics assessments},
journal = {Teaching Statistics},
volume = {48},
number = {1},
pages = {19-32},
doi = {https://doi.org/10.1111/test.70013},
year = {2026}
}

@article{safilian2025ratas,
  title={Ratas framework: A comprehensive genai-based approach to rubric-based marking of real-world textual exams},
  author={Safilian, Masoud and Beheshti, Amin and Elbourn, Stephen},
  journal={arXiv preprint arXiv:2505.23818},
  year={2025}
}

@article{ilieva2025framework,
  title={A {F}ramework for {G}enerative {AI}-{D}riven {A}ssessment in {H}igher {E}ducation},
  author={Ilieva, Galina and Yankova, Tania and Ruseva, Margarita and Kabaivanov, Stanimir},
  journal={Information},
  volume={16},
  number={6},
  pages={472},
  year={2025},
  publisher={MDPI},
  DOI = {10.3390/info16060472}
}

@article{pedregosa2011scikit,
  title={Scikit-learn: Machine learning in Python},
  author={Pedregosa, Fabian and Varoquaux, Ga{\"e}l and Gramfort, Alexandre and Michel, Vincent and Thirion, Bertrand and Grisel, Olivier and Blondel, Mathieu and Prettenhofer, Peter and Weiss, Ron and Dubourg, Vincent and others},
  journal={Journal of Machine Learning Research},
  volume={12},
  pages={2825--2830},
  year={2011},
  publisher={JMLR. org}
}

@article{carscadden22,
  title={To Tidy or Not When Teaching R Skills in Biology Classes.},
  author={Carscadden, Kelly and Martin, Andrew},
  journal={International Journal of Higher Education},
  volume={11},
  number={5},
  pages={39--50},
  year={2022},
  publisher={ERIC}
}

@article{cetinkaya21,
   title={An educator’s perspective of the tidyverse},
   volume={14},
   DOI={10.5070/t514154352},
   number={1},
   journal={Technology Innovations in Statistics Education},
   publisher={California Digital Library (CDL)},
   author={Cetinkaya-Rundel, Mine and Hardin, Johanna and Baumer, Ben and McNamara, Amelia and Horton, Nicholas and Rundel, Colin},
   year={2022}
}

\end{document}